# Adoption of Industry 4.0 Technologies in Airports
## - A Systematic Literature Review

Jia Hao Tan and Tariq Masood
University of Cambridge, Changi Airport Group, and University of Strathclyde

## Pre-Print

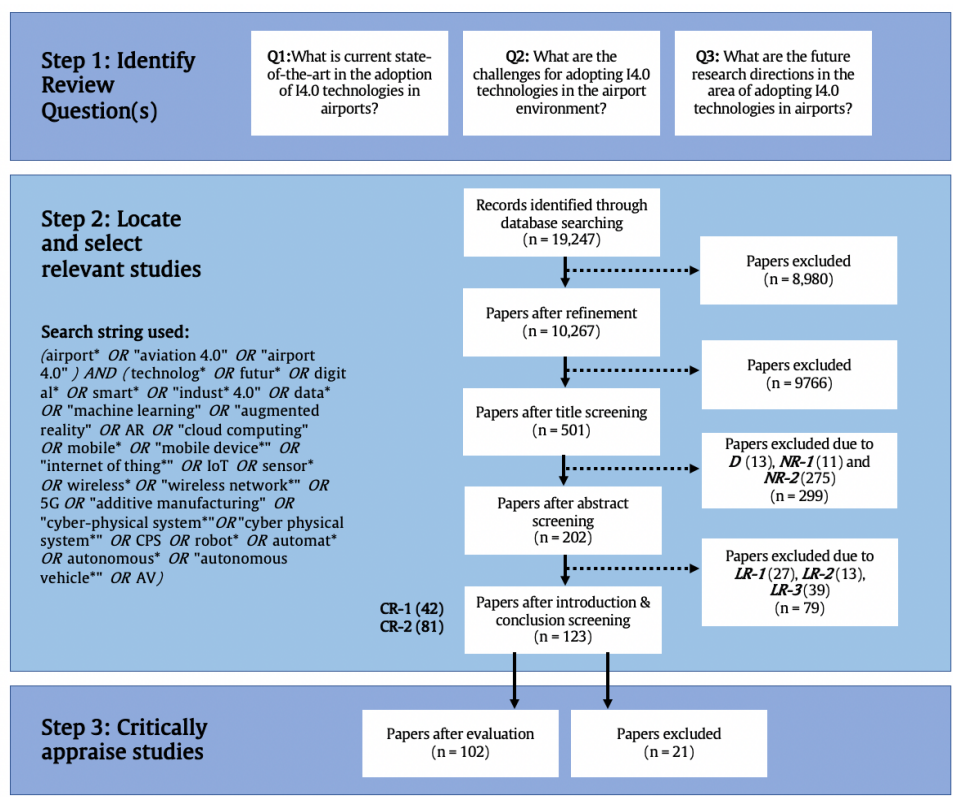

## Highlights

o Airports have been constantly evolving and adopting digital technologies to improve operational efficiency, enhance passenger experience, generate ancillary revenues and boost capacity from existing infrastructure.
o The COVID-19 pandemic has also challenged airports and aviation stakeholders alike to adapt and manage new operational challenges such as facilitating a contactless travel experience and ensuring business continuity.
o Digitalisation using Industry 4.0 technologies offers opportunities for airports to address short-term challenges associated with the COVID-19 pandemic while also preparing for future long-term challenges that ensue the crisis.
o Through a systematic literature review of 102 relevant articles, this article discusses the current state of adoption of Industry 4.0 technologies in airports, the associated challenges as well as future research directions.
o The results of this review suggest that the implementation of Industry 4.0 technologies is slowly gaining traction within the airport environment, and shall continue to remain relevant in the digital transformation journeys in developing future airports.

# Title: Adoption of Industry 4.0 technologies in airports - A systematic literature review

**Authors:** Jia Hao Tan[1,2]*†, Tariq Masood[1,3]*†.

**Affiliations:**

[1]Institute for Manufacturing, Department of Engineering, University of Cambridge, 17 Charles Baggage Road, Cambridge, CB3 0FS, UK.

[2]Changi Airport Group (Singapore) Pte Ltd, 60 Airport Boulevard, Changi Airport Terminal 2, Singapore 819643.

[3]Department of Design, Manufacturing and Engineering Management, University of Strathclyde, 75 Montrose Street, Glasgow, G1 1XJ.

*Correspondence to: jht42@cantab.ac.uk (Jia Hao Tan); tariq.masood@strath.ac.uk (Tariq Masood)

†Equal contributions

**Abstract:**
Airports have been constantly evolving and adopting digital technologies to improve operational efficiency, enhance passenger experience, generate ancillary revenues and boost capacity from existing infrastructure. The COVID-19 pandemic has also challenged airports and aviation stakeholders alike to adapt and manage new operational challenges such as facilitating a contactless travel experience and ensuring business continuity. Digitalisation using Industry 4.0 technologies offers opportunities for airports to address short-term challenges associated with the COVID-19 pandemic while also preparing for future long-term challenges that ensue the crisis. Through a systematic literature review of 102 relevant articles, we discuss the current state of adoption of Industry 4.0 technologies in airports, the associated challenges as well as future research directions. The results of this review suggest that the implementation of Industry 4.0 technologies is slowly gaining traction within the airport environment, and shall continue to remain relevant in the digital transformation journeys in developing future airports.

**One Sentence Summary:** Industry 4.0 technologies have potential to revolutionize the future of airports.

## 1. Introduction

Airports are the first touchpoints for passengers whenever they transit or arrive in a destination country by flight. With the expected increase in passenger and cargo traffic in the longer term, airports of the future will face the challenge of delivering the same level of service excellence with increasingly strained resources. These service parameters include, but are not limited to flight punctuality rate, ground handling competence, air transportation safety, etc (*1*). Furthermore, the ongoing COVID-19 pandemic is also changing the way airports operate in terms of new health and safety guidelines such as introducing safe distancing measures and enhanced cleaning and disinfection of terminals (*2*).

Technology offers opportunities for airports and aviation stakeholders to address immediate operational challenges associated with the COVID-19 pandemic while preparing for a post-COVID-19 world. Apart from capacity expansion efforts, airports have also taken the route of digitalisation to achieve the objectives of improving operational efficiency, enhancing passenger experience, generating ancillary revenue and boosting capacity from existing infrastructure (*3*). The technologies adopted such as wireless sensor networks (WSN) to monitor airfield lightning systems (*4*) and service robots for wayfinding (*5*) are closely associated with Industry 4.0 (I4.0) technologies used in manufacturing systems (*6*). This has led to the development of related alias such as Airport 4.0 to describe the next generation of digitally transformed airports (*3*).

This review examines the state-of-the-art in the adoption of I4.0 technologies in airports through a systematic literature review, and identifies associated challenges and future research directions in this field.

The rest of this article is structured as follows. Section 2 provides background of I4.0 technologies. Structured literature review (SLR) methodology is presented in section 3. Based on the SLR results, section 4 discusses the state of the art of Airport 4.0 before analysing the challenges in adopting I4.0 technologies in airports in section 5. Section 6 charts the way forward for Airport 4.0.

## 2. Industry 4.0 technologies

"Industrie 4.0" (or Industry 4.0 or I4.0) was first introduced by the German government in 2011 as part of the "High-Tech Strategy 2020 Action Plan" (*7*). I4.0 signifies the advent of the fourth industrial revolution which involves an overhaul of traditional manufacturing environments by incorporating advanced digitalization (*8*). This transformation is characterized by the development and proliferation of Cyber-Physical Systems (CPS) which connects physical operations with computing and communication infrastructures through networking, thereby enabling manufacturing and service innovations (*9*, *10*).

By nature of its definition, I4.0 is commonly associated with the manufacturing industry. The term "smart factories" is used to refer to factories with intelligent CPS that produces material goods in a highly mechanised and automated manner, thereby achieving high levels of productivity and operational efficiencies (*11*). I4.0 technologies have revolutionised traditional manufacturing environments by achieving flexibility and higher automation such as through the use of virtual reality (VR) for the design and simulation of a human-robot production system (*12*), and for future-proofing against future disruptions and changes(*13*). More recently, these I4.0 applications have brought about capacity and efficiency improvements to meet the short-term demand for medical equipment during the COVID-19 pandemic (*14*), and allow for the modelling of population movements and the spread of diseases using big data (*15*, *16*).

Aside from manufacturing applications, I4.0 technologies can be applied in service industries as well, which was consistent with the concept of servitisation where both service and manufacturing companies alike can provide customer-focused collection of goods, services, support, self-service and knowledge to establish a competitive advantage among other players (*17*) .

There is no consensus in literature in defining the technologies that make up I4.0 (*18*), including variations in the number of categories that I4.0 technologies fall into (*6*, *19*). To reduce ambiguity, this article will classify the I4.0 technologies into seven categories based on technology functionalities (*18*) as seen in Table 1. The functional classification was preferred due to the interdisciplinary nature of the I4.0 concepts and technologies which makes it difficult to provide clear distinctions among them (*18*).

## Table 1: Key Industry 4.0 functional categories and their related technologies

| Functional category | Description | Related technologies |
|---|---|---|
| Data analysis & processing | Technologies used for information processing. | Machine learning, data mining, artificial intelligence, authentication, blockchain |
| Simulation, visualisation & modelling | Technologies used for increased perception, visualisation and utilisation of information. | Augmented reality (AR), virtual reality (VR), simulation, digital twin, building information modelling (BIM) |
| Cloud computing | Technologies to enable the delivery of computing services over the Internet. | Cloud storage databases, servers, cloud-based collaboration tools |
| Mobile smart devices | Technologies that are portable and makes use of mobile terminals for accessing information. | Smartphones for accessing digital channels, tablets, smart glasses, wearables |
| Internet of Things (IoT) | Technologies that make use of intelligent devices and sensors for communication and presentation of information. | Sensors such as RFID, LIDAR, etc, wireless sensor networks (WSN), location detection technology, 5G networks |
| Additive manufacturing | 3D printing technology for product development and customisation of goods on a large scale. | 3D printing |
| Cyber-physical systems (CPS) | Systems or networks that make use of autonomous elements involving human-machine and machine-machine interfaces to coordinate and automate processes. | Robots, automatic guided vehicles (AGVs)/ automated vehicles (AVs), self-service technologies (SSTs) |

## 3. Methodology

To understand the current state of adoption of I4.0 technologies in airports and the associated challenges, a systematic literature review (SLR) was conducted using a 3-step methodology adapted from frameworks proposed in literature (*20, 21*). The overall methodology for the SLR and the search string used are illustrated along with the key outcomes of each stage (Fig. 1).

After the 3-step process, 102 papers were identified to be evaluated in full. These papers have clearly explained research methodologies, research results and are relevant to the research topic. Table 2 shows a representative part of the document used to capture information from the papers reviewed.

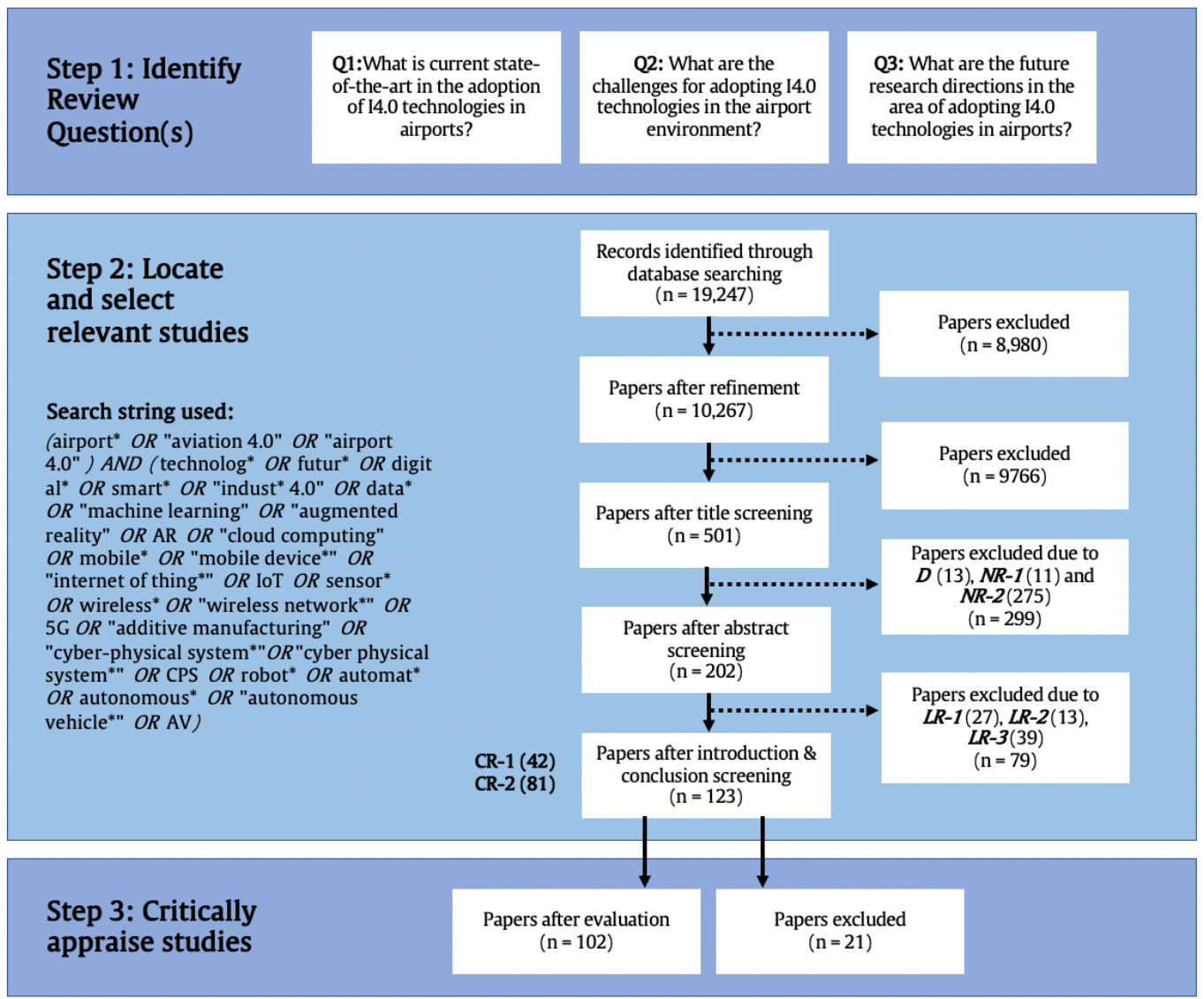

**Fig. 1. Overall methodology for systematic literature review.**

## 4. Airport 4.0

Establishing clear definitions provide the fundamental starting ground for research. Out of the papers reviewed, only four papers mentioned the term "Airport 4.0" in the title, abstract, list of keywords or the general content (Table 2). The authors discussed applications involving data analytics and sensor technologies (*22, 23*) and emphasised data analytics as a core capability of Airport 4.0 by illustrating a range of technologies that support the objectives of improving efficiency of airport operations, enhancing passenger journey and creating ancillary revenues (*3*). From these papers, it is evident that the term Airport 4.0 is used as a broad term to imply advancement and adoption of technologies in airports, with a weak reference to the associated term "Industry 4.0".

Despite a lack of academic publications that mention Airport 4.0, I4.0 technologies have increasingly been adopted in airports around the world across various airport functions.

**Table 2 (part 1). Review of literature on Industry 4.0 technologies in airports.** Representative sample of key information captured from reviewed papers.

| Authors | Year | Description of application | Technology | Application area | Challenges in adoption | Future research directions |
|---|---|---|---|---|---|---|
| Lambelho et al. | 2020 | Using machine learning to predict whether strategic scheduled flights will be delayed/ cancelled | Data analytics (machine learning) | Air traffic control and management | Accuracy of prediction (limited by features selected) | Improve accuracy by including more features into prediction model. Generate strategic insights (of impact from cancellations) |
| Zhang and Xue | 2019 | Using Data mining methods (Association, clustering) to identify areas of improvement in airport to assess comprehensive airport ground support capability | Data analytics (data mining) | Airside operations | Need for customisation (Results are specific to the particular airport depending on data) | Apply similar method on data of other airports |
| Huang et al. | 2019 | Using Data (wifi positioning data) to portray passenger behaviour in terminals, model and predict occupancy/ dwelling time in terminals to manage energy consumption | Data analytics, sensor technologies, IoT | Airport terminal operations | Data limitation (Wifi positioning data is still limited in resolution and vulnerable to noise). Need for customisation (Parameters needs to be recalibrated for different terminal features) | Expand scope of application |
| Negri et al. | 2019 | Biometric technology for airport self-service check-in | Biometric technology, SST | Airport terminal operations | High initial investment, privacy concerns, Need for accompanying technology/ processes, cooperation of multiple stakeholders), need for usability tests | Expand perception study of passengers towards biometric SSTs to other airports |
| Kovynyov and Mikut | 2019 | Digital technologies in airport ground operations | SST, AVs, Biometrics, RFID, IoT | Airport terminal operations, Airside operations, Airport logistics | Technology readiness (Some technologies belong to low technology readiness levels), so continued trials are necessary | Quantitative criteria for evaluation of technologies in airports, Incorporate uncertainty into forecasting and planning, Multi-objective optimisation, Create new business models from collected data |
| Groskreutz et al. | 2019 | Using big data analytics (data from ADS-B antenna) to determine taxi congestion locations | Data analytics | Airside operations | Data storage (cumbersome, expensive, time-consuming) | Expand scope of application (runway occupancy time analysis, rapid exitway prediction, etc) |
| Cho and Kwag | 2019 | Using big data analytics for flight demand forecast for terminal planning | Data analytics | Airport planning, design and construction | Data limitations (Forecasts are dependent on assumptions made in parameters used) | Include more parameters in model |
| Di Vaio and Varriale | 2019 | Blockchain technology for integration of information and data for air traffic management | Blockchain | Airside operations | Technology disruptions, Need for relevant skills | Expand scope of analysis to include more airports. |
| Thomopoulo et al. | 2019 | Deep learning for risk-based security assessment of passengers | Deep learning, data analytics, simulation | Airport security | Compliance with privacy standards, Need for continued trials (validation of performance of technology when adopted), Financial investment, technology limitations (lack of sufficient data, assumptions used, determining the right parameters) | Include more parameters in model. Use real-world data |
| Conversy et al. | 2018 | Cyber-physical system involving the use of a domain-specific graphical language to author and operate air traffic control operations | CPS, human-machine interface | Air traffic control and management | Need for customisation (Results are specific to the particular airport depending on data); Technology limitations (accidental actions triggered by humans/ high impact decisions made by automation), Need for specialised skills and training (May not have proficiencies to program automation in ATCO) | Further tests and trials. Expand scope of application to other airports |
| Hättenschwiler et al. | 2018 | Explosive detection systems for cabin baggage screening (EDSCB) as a diagnostic tool for airport security X-ray screening and/or automatic decision | Data analytics (AI) | Airport security | Technology limitation (high false alarms), User receptivity (cry wolf effect for experienced screeners) | Further tests in an operational environment |

**Table 2 (part 2). Review of "Airport 4.0" literature.**

| Authors | Year | Title | Description | Research Methodology | Reference to Airport 4.0 |
|---|---|---|---|---|---|
| Büyüközkan et al. | 2020 | Analysis of success factors in aviation 4.0 using integrated intuitionistic fuzzy MCDM methods | Proposed methodology to prioritise success factors of aviation 4.0 among 3 airline companies using MCDM methods | Literature review | Moderate. Airport 4.0 and Aviation 4.0 used to describe the era of digital transformation in line with Industry 4.0 |
| Koenig et al. | 2019 | Innovative airport 4.0 condition-based maintenance system for baggage handling DCV systems | Pilot trial in Heathrow airport to collect real-time high frequency vibration data for baggage carts to facilitate predictive/condition-based maintenance of baggage handling system. | Case Study | Limited. Airport 4.0 is only mentioned in title. |
| Zaharia and Pie | 2018 | Challenges in airport digital transformation | Challenges in adopting digital solutions adopted across different areas in Henri Coanda International Airport | Case Study | Moderate. Airport 4.0 used as a section header with no clear defintion and as a diagram label with a suite of technologies to achieve operational efficiency, passenger journey and retail ancillary revenue. |
| Felkel et al. | 2017 | Hub airport 4.0 - How frankfurt airport uses predictive analytics to enhance customer experience and drive operational excellence | Overview of recent projects in Frankfurt airport that made use of predictive analytics to improve passenger flow management, optimising aircraft positioning and forecast retail revenue. | Case Study | Limited. Airport 4.0 is only mentioned in title. |

*4.1 Passenger handling services*

The use of mobile devices (iPad) has greatly enhanced customer service by providing a personalized touch for each passenger. Airport agents have begun using digital tablets to provide real-time access to passenger information in order to better assist them with their queries (e.g. flight time, baggage information, etc) (*24*). Mobile devices can also be incorporated in a wider IoT-based smart airport solution where passengers and airport employees can access digital channels or airport mobile applications to track luggage, manage waiting time, car parking, etc (*25*). This smart airport solution involves an extensive IoT network with multiple sensors (e.g. line queue sensor, car parking sensor, etc) to collect data, the use of AI to analyse data and respond accordingly, as well as the use of cloud services to facilitate data management and application processing (*25*, *26*). Passengers' mobile devices can also be a platform to crowdsource data to identify delays and points of congestions within the terminal, and provide insights on how the allocation of resources can be better optimized. For example, data from Bluetooth-enabled mobile devices at George Bush Intercontinental Airport was used to identify security operations as a source of congestion (*27*). A model was also developed to predict occupancy and dwelling time of passengers in Shanghai Hongqiao International Airport based on Wi-Fi indoor positioning data (*28*).

Social robots are used as a method to assist passengers with wayfinding in airports (*29*). The SPENCER project was launched in the European Union (EU) to develop a mobile robotic platform to guide passengers at Amsterdam Schiphol airport from their arrival gates to passport control. The use of Inverse Reinforcement Learning enabled the SPENCER robot to be socially aware by behaving in accordance to social norms when approaching passengers (*30*). Aside from social robots, autonomous cleaning robots have also been developed and used for the cleaning of airport terminals and terminal exteriors (*31*).

Cloud computing also facilitates the implementation of Self-Service Technologies (SSTs) in airports such as the common use self-service (CUSS) kiosks for check-in as well as airport retail or information kiosks for product search, wayfinding and gift registry, etc (*32*). Overall, SSTs facilitates seamless passenger processing in airports and can potentially reduce the operating costs for airports by 95.6% per passenger as compared to standard check-in (*33*, *34*). Furthermore, a survey study has revealed that SSTs contribute to improved passenger satisfaction by allowing passengers to feel more confident about technologies that allow them to retain control and independence (*33*).

*4.2 Commercial services*

Commercial activities within the airport make up the non-aeronautical revenue stream of airports, which can take up as much as 45 percent of the airport's total revenue such as in the case of Asia-Pacific airports on average (*35*, *36*). These activities include retail, food-and-beverages (F&B) and duty-free concessionaires, car-parking and leasing of office spaces (*35*). Frankfurt airport's Smart Data Lab made use of predictive analytics to forecast retail revenue based on passenger count and time, and also investigate how aircraft positioning plans can be optimized to influence shopping behaviour and increase retail revenue (*22*). This provides opportunities for the airport to explore business opportunities for the exclusive or prioritized use of certain terminal areas identified as "prime" retail spots (*37*). The use of data analytics on passenger data enables the profiling of passenger attributes such as demographics and behaviours. These profiles can be further segmented and used for targeted marketing by airport concessionaires to push retail advertisements to passengers' mobile devices (*38*).

Technology and digital platforms also show prospect for creating *airportainment* services to enhance passenger experience and improve retail revenue, such as the use of virtual shopping platforms and smart sensory panels for advertising (*39*). In 2013, Changi Airport Group launched *iShopChangi*, an e-commerce platform that allows passengers to shop for duty-free products at the convenience of Internet-enabled devices, thus capturing a wider target audience such as the increasingly tech-savvy millennials (*40, 41*). Blockchain technology enhances the security of the personal data provided by passengers, which may encourage passengers to share more data in exchange for personalized recommendations of products and services. The use of AI in developing chatbots allow passengers to receive recommendations of products and services at different stages of travel, including on e-commerce platforms for airport retail (*42*).

### *4.3 Airside ground operations*

Airside ground operations refer to services that are required by airlines between landing and departure of the aircraft (*43*). The airside environment contains assets which need to be regularly maintained to ensure operational safety. I4.0 technologies help to improve the efficiency of operations within the airside environment.

Cameras and sensors placed in the airside environment (e.g. transponders, antenna, array, radar, light detection and ranging (LIDAR), and piezoelectric sensor network) can be used to replace visual inspections in airport surface area surveillance (ASAS) for maintenance, the monitoring and detection of foreign object debris (FOD), aircraft and human movements, early fault detection, etc (*44–49*). When coupled with artificial intelligence and video surveillance, airside operations can be closely monitored to ensure an efficient and safer operating environment in the apron. For example, the use of video data and clustering algorithms allow the surveillance system to detect the presence and absence of wheel chocks for specific aircrafts parked in the apron and raise alerts when necessary (*50*). The real time sensor data obtained is useful for maintenance such as in monitoring the structural performance of airport asphalt pavement using Fiber Bragg Grating (FBG) measurement technology and providing possible avenues for predictive maintenance (*51, 52*).

Analysing available data from the airside can also improve the efficiency of airside operations. Based on data collected regarding weather and flight operations, Frankfurt airport's Smart Data Lab used predictive analytics to predict the arrival time of aircrafts at the parking stand. This facilitated on-time services by ground handlers to manage aircraft turnaround activities (e.g. baggage unloading, freight unloading, etc) which reduced the waiting time for passengers and also reduced the cost of ground handlers due to idle capacity (*37*). Furthermore, the data collected in the airside can be further processed using algorithms to optimize airside operations such as improving the ecological impact of de-icing practices in airports, minimizing taxiing time and fuel consumption of taxiing aircraft, etc (*53, 54*). Airside data can also be used to develop metrics such as runway utilization, average taxi-out times and departure spacing efficiency to assess and quantify airport performance, thereby generating insights for improvements (*55*).

Autonomous robotic vehicles provide an alternative to human labour for ground operations, including aircraft-towing, cargo-handling and passenger transport (*56*). These autonomous robot applications can also be extended to involve multiple robots. A JAVA simulation of a multi-robot control system involving a LadderBot (ladder for passenger boarding), LuggageBot (for luggage delivery to plane) and PassengerBot (to transport passengers from terminal to plane) was carried out to assign tasks to the available robots based on a set of criteria (*57*). The simulation results show more than 70% reduction in total operation time from the collective action of multiple robots as compared to a single robot operation. Germany's Frankfurt Airport trialled the use of taxiing robots ("TaxiBot") for the autonomous taxiing of aircrafts from the apron to the runway. These TaxiBots incorporate sensors to detect obstacles such as airplanes, people and vehicles to prevent collisions (*58*).

### *4.4 Air traffic control and management*

Air traffic control and management is traditional reliant on humans due to safety concerns. Given the increasing air traffic and operational complexity, air traffic control and management can benefit greatly from the use of automation alongside humans by corroborating with the judgement of air traffic controllers (ATCOs) (*59*, *60*). For example, the use of a neural network as a nowcasting tool based on sonic detection and ranging (SODAR) data allows for the forecast of low wind profiles to help ATCOs make rapid and correct decisions on runway utilisation while considering air traffic flux optimisation and safety requirements (*60*). Data mining methods were used on historical datasets regarding fog observations at Paris Charles de Gaulle airport to develop a fog nowcasting model to allow air traffic controllers to mitigate adverse visibility conditions in advance (*61*). A winter weather forecasting system WHITE (Winter Hazards in Terminal Environment) was developed using meteorological data to nowcast and estimate the beginning and duration of hazardous winter weather conditions in Munich Airport to facilitate decision making by traffic controllers. Furthermore, WHITE integrated a participatory sensing approach involving the use of high-capacity sensor-equipped mobile phones as ubiquitous sensing devices of meteorological parameters (*62*).

Data analytics and machine learning also allow for the development of models to predict departure capacities, taxi-out times, arrival/departure delays and runway configuration decisions considering different influencing factors such as weather, etc. This information can improve flight planning in airports to reduce potential flight delays and passenger backlogs (*38*, *63–66*). Aside from providing real-time information for air traffic control, the availability of data from ground-traffic surveillance system also allow airport operators to determine quantitatively if the risk of runway overrun in the airport is valid and mitigate the risks if necessary (*67*).

The use of ADS-B data which determines an aircraft's position via satellite navigation allowed researchers to reconstruct the taxiing route of aircrafts in Guangzhou Baiyun airport and identify environmental factors that cause taxiing delays. These allowed for the construction of models to optimize aircraft taxiing schemes in the airport to reduce taxiing delays, which are a major cause of flight delays (*68*). By combining ADS-B data with video monitoring data using data fusion methods, aircrafts in apron can be automatically labelled for better monitoring and control of ground traffic (*69*).

### *4.5 Airport logistics*

Airport logistics involves the movement of goods and includes activities such as cargo handling, baggage handling and catering handling.

The integration of radio-frequency identification (RFID) and WSN allows for the establishment of a cargo monitoring system that can achieve real-time tracking and positioning of airport cargo to facilitate smooth operations (*70*). RFID can also be used for the tagging of luggage with the information collected being stored in an IOT cloud server for easy retrieval of data at different airports. When integrated with mobile applications, passengers can use their mobile devices to accurately track the location of their luggage and reduce instances of lost luggage (*71*, *72*). A similar alternative to luggage tracking used a low power luggage localization system which involves a battery-operated smart tag that sends real-time tracking information to the passenger's mobile application or smart wearable (*73*).

The transport of baggage in airports are facilitated by baggage handling systems (BHS) which transport baggage on carts at high speeds. Heathrow Airport carried out a pilot trial using IOT sensors and data analytics for predictive maintenance of baggage carts in the BHS (*23*). The use of mobile devices can also improve the efficiency and ease of BHS maintenance operations. Four activities relating to BHS maintenance (arrival to site, utilization of field manuals, preparation of test sheet and preparation of error report) were identified that can be incorporated into mobile devices (*74*).

### *4.6 Airport planning, design and construction*

As the aviation industry adapts to the evolving needs of passengers in the 21st century (post COVID-19), it adds to the already complex and fragmented systems within the airport environment. Airport infrastructure is highly varied and complex, including but are not limited to terminals, apron, runways, carparks, roads, air traffic control tower, etc. The assets and infrastructure networks in an airport needs to be carefully designed in order to meet the operational requirements of the airport (*75*, *76*).

Building Information Modelling (BIM) provides a digital 3D representation of the physical and functional characteristics of the facility. It is instrumental as an I4.0 technology to facilitate the collaboration and virtual designing of airport infrastructure by all project stakeholders, thereby achieving greater efficiency and productivity in project management (*77*). Aside from visualization, BIM integrates the digital model with information to allow for the simulation of operations and building models, as well as program optimization and spatial design optimization (*78*). An example is the use of BIM in modelling engineering systems within the terminal building, piers, air traffic control (ATC), runways and sitewide facilities for Istanbul New Airport (INA), which is an international airport in Turkey that was claimed to be the largest airport in the world when completed (Gao, 2018). In the INA project, mobile devices (iPads) were also used during airport construction, allowing ready access to cloud-based collaboration tools (Autodesk BIM 360 field, BIM 360 Glue) for project management, quality control and assurance management, etc (*75*).

Simulation technology can create a virtual environment that can facilitate airport planning and design. The use of gaming technology was explored to develop a multi-agent crowd simulation model of a terminal in London Gatwick airport considering unique traits of airports such as time pressures, travelling purposes and heightened emotions (*79*). The simulation was able to generate believable and intelligent crowds at an interactive frame-rate.

The availability of ground traffic surveillance systems data can also be used for airport master planning to assess the capacity of the airport. The data can be analysed to better understand runway traffic and derive the relevant factors that affect runway throughputs (*67*). Based on historical air traffic data, an airport based queuing network model was developed to simulate delay propagation across the European Air Traffic Network and derive the "breaking point" of an airport (*80*), which is defined as the critical capacity value of the airport below which there is a sudden sharp increase in total delay. Furthermore, the available data can be used in computer simulation to allow airport planners to better predict the future tendency of the airport system. The use of simulation is prevalent in airports to simulate a range of applications in the airside (flight delays, runway systems, taxiway systems, etc) and the terminal (passenger flow, baggage handling, etc) (*81*). For example, a simulation model was developed based on current traffic data and weather conditions of Lelystad airport to determine future capacity and performance of the airport (*82*). Such insights are useful for the master-planning of airports to inform future airport expansion or construction decisions in the design and analysis phases.

### *4.7 Airport security*

Since the 9/11 attacks in New York and Washington in 2001, which registered 2,938 deaths, governments around the world have stepped up security measures and systems in airports. Along with the hiring and training of security personnel, heavy investment in technology has also played a significant role in preventing a repeat of similar terror attacks against aviation (*83*).

Explosive detection systems for cabin baggage screening (EDCSB) make use of X-ray technology as well as artificial intelligence to detect and discern explosives from X-ray images of baggage. These EDCSB systems can automatically decide if the baggage contains a threat or be used as a diagnostic tool that raises alarms for further inspection by airport security personnel (*84*, *85*). Automated Target Recognition (ATR) software within millimetre-wave body scanner machines allows for the automatic review and analysis of images for concealed threats (*86*).

The use of facial recognition technology and biometric data has facilitated automatic border control e-gates, which are example of SSTs that allow passengers to pass through immigration checks without direct service employee involvement (*87*, *88*). Blockchain technology can provide a trusted network for the storage of biometric and other personal data of passengers, while providing a platform for collaboration between various airport stakeholders. These data can facilitate faster passenger journeys while mitigating cybersecurity and privacy concerns (*42*). Airport surveillance video data can also be stored on cloud computing platforms for easy retrieval to detect face images and determine the specific identity of individuals (*89*).

Perimeter security is another important aspect of airport security. A Perimeter Intrusion Detection System (PIDS) was developed based on WSN involving the use of a network of self-sustaining accelerometer-based sensor nodes which will detect intrusion events and prompt automatic responses from the signal base (*90*). Furthermore, new technology involving the software coupling of the radar sensor and cameras in an integrated radar-camera system allow the cameras to automatically track the targets detected by the sensors (*91*). A pattern recognition method based on neural network technology also improves the intrusion detection probability of airport perimeter security systems by determining whether an intrusion has occurred as well as the type of intrusion, which reduces the risk of false alarms (*92*).

The terminals and gate areas within an airport are also closely monitored for suspicious personnel and hazardous materials. The EU-funded ATOM project that developed a multi-sensor approach integrates active and passive radar sensors with currently used surveillance systems, fuses data obtained from the sensors and triggers an automatic alert to security operators to neutralize the threat (*93*). The data fusion process of airport surveillance data can be improved using multi-sensor data fusion algorithms and machine learning to estimate target characteristics accurately in the presence of dense traffic and environmental noise clutter in airports (*94*).

The use of data analytics also allow airports to forecast and plan for resources required in security screening processes (e.g. number of staff required, number of screening lanes, etc) to meet expected demand, evaluate performance and refine the planning process (*95*).

## 5. Challenges in adopting I4.0 technologies in airports

While there is evidence of I4.0 applications in the airport, the SLR has also revealed technological, operational and environmental challenges that limit the widespread adoption of I4.0 technologies in the airports of the future (Fig. 2).

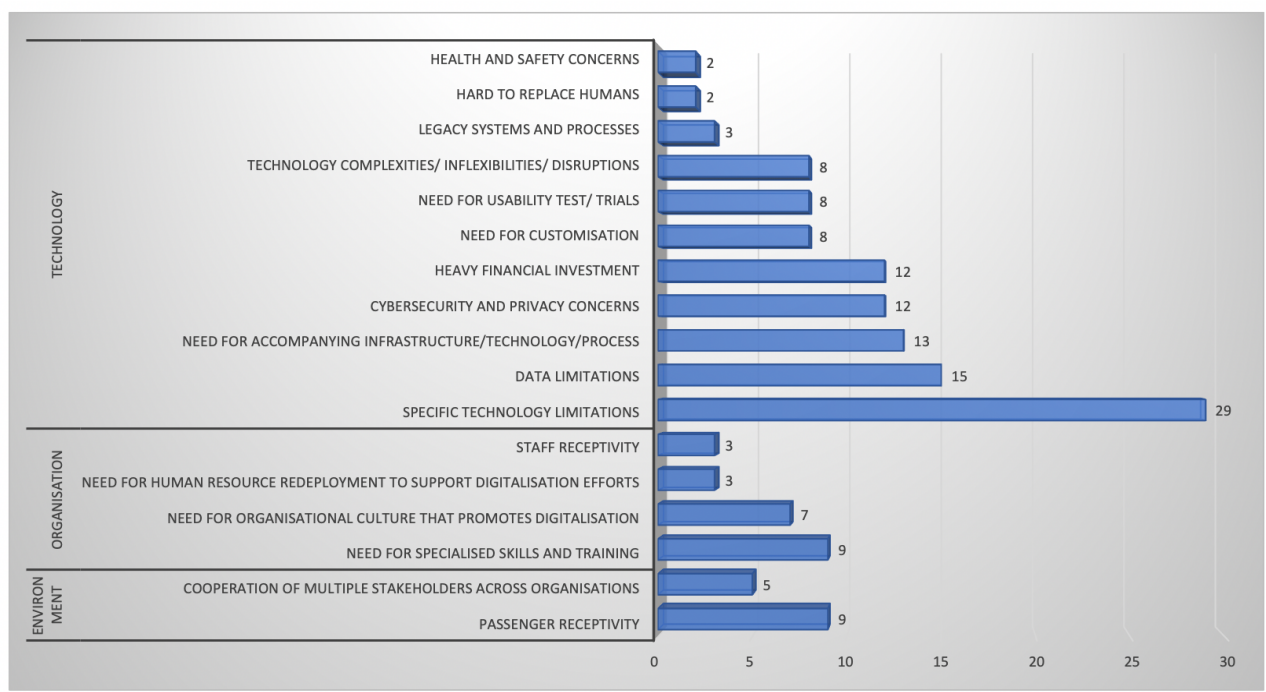

**Fig. 2. Technological, organizational and environmental challenges for adopting I4.0 technologies in airports.** Bars represent number of publications.

### *5.1 Technological challenges*

The adoption of I4.0 technologies is subjected to specific technology limitations, cybersecurity and privacy concerns. For example, the feasibility of battery-operated devices such as a battery-operated smart luggage tag depends heavily on the power consumption of the device and how long the battery can last before a recharge is necessary (*73*). The use of LIDAR sensors for the detection of Foreign Object Debris (FOD) detection in the apron is promising for large and mid-size standardized objects, but is not equally effective in detecting extremely small FOD or objects with significant reflection properties (*47*). A stress-test of the automated border control e-gate at Adolfo Suárez Madrid-Barajas Airport in Spain was carried out by subjecting the system to intentional presentation attacks to cheat the verification system involved (*88*). The results showed that the facial recognition technology involved remains vulnerable to presentation attacks such as the use of the fake photos and morphing attacks (blending the face of the traveller and attacker using image fusion software). These security vulnerabilities not only reflect the limitations of the technology but highlights the importance of introducing supporting technology such as increased surveillance to complement its use. Furthermore, when adopting new security screening technologies, passengers were less receptive of new screening processes that invade their privacy more than what is currently practised (*96*).

The lack of accurate, complete and rich data limit the effectiveness of data analytics in producing descriptive, predictive and prescriptive models to improve airport operations (*67*). The lack of good quality data also reduces trust towards the insights obtained from data analytics (*95*). Furthermore, in some cases, additional data needs to be obtained to complement existing data to support deeper analysis. For example, the availability of smart card data at Pudong International Airport allowed operators to understand airport parking patterns, but additional information regarding socio-economic data and individual parking activity of passengers is required in order to forecast parking demand (*97*).

The successful implementation of certain I4.0 technologies require accompanying infrastructure, technology or processes which may not be available in the airport. The importance of supporting technologies and infrastructure which are not directly focused on service optimization processes can be essential to complement and facilitate core airport technologies (*33*). For example, with the introduction of digital platforms to encourage the use of mobile devices for wayfinding and the retrieval of baggage information, the presence of guest-related amenities such as universal battery chargers in guestrooms and the availability of reliable Wi-Fi infrastructure are essential to improve the overall passenger experience (*33*, *71*, *98*). The presence of established systems or processes may also impede the implementation of new or advanced technologies in a smart airport environment. For example, the presence of standardized construction contracts which prevent the use of innovative BIM technology for on-site operations for airport construction (*99*). Furthermore, the use of non e-passports that do not contain biometric information also prevent passengers from using automatic border control gates in airports (*87*).

As airports worldwide continue to invest in smart technologies, IT expenditure by airports has increased considerably and is expected to top $4.6 billion by 2023 (*42*). Even though digital transformation of airports provide growth opportunities for airports to modernize their infrastructure, overcome physical infrastructure constraints and enhance the overall airport experience, the significant costs of implementing digital technologies require airports to carry out a cost-benefit analysis to justify the spending on smart technologies (*3*, *100*). Furthermore, airports vary in terms of design and operational aspects, thus some technology solutions need to be customised to the specific airport context in order to improve performance. Airport end-users need to be involved in the development of technology applications in order to design applications that are catered to the needs, desires and fears of end-users (*87*). This involvement stretches from initial pilot tests during development to post-implementation testing with real users in order to further enhance the system or technology application (*87*, *96*). This contributes to significant time and resources that needs to be invested to develop minimum viable product (MVP) applications, collecting user feedback and iterating the design of the technology application (*101*).

The challenge of involving humans in the implementation of new technology is particularly evident in situations where the technology adopted cannot fully replace humans, or still require a human operator such as in air traffic control and automated border control. It is challenging to improve human-automation collaboration, specifically in the context of air traffic control operations, by considering the human agent and the automated environment as a joint cognitive system without compromising on performance (*102*). Health and safety concerns also affect the feasibility of implementing I4.0 technologies in the airport environment. Safety is the main challenge towards the successful deployment of autonomous robots in airports, despite the operational benefits that these robots present (*103*). These safety concerns are particularly valid in the dynamic airside environment where robots may pose as a hazard to other moving operators and service vehicles.

## *5.2 Organisational challenges*

Airport operators need to build an organisational culture that promotes digitalisation and ensure that the entire company is aligned to a common goal towards digitalisation. This involves fostering acceptance of the benefits of technology and persuading stakeholders to share their data for further processing, and to verify the relevance and feasibility of execution of any ideas (*3, 37*). Airport operators also need to anticipate and prepare their workforce for the demands of digital transformation by introducing and promoting relevant training and education programmes (*3, 104*). When technology is used to complement human efforts, user mentality and receptivity may limit the effectiveness of the technology application. For example, the 'cry-wolf' effect was observed where experienced airport security officers ignore the warnings from cabin baggage screening machines which serve as diagnostic aids for explosives (*84*). This effect is more pronounced in the real world when the frequency of threats is low and where most warnings from EDSCB machines are false alarms. It was also a challenge to convince managers to use data processing and analytics to make smarter decisions, particularly if the decision-making is traditionally done using an experience or command-driven approach (*37*).

Furthermore, employees need to be dedicated towards working on innovation projects (*37*). For the case of the Smart Data Lab project in Frankfurt Airport, employees had to be taken off their day jobs for an intense huddle period to generate insights from data. This implies the challenge for employees to partake in digital transformation efforts while being involved in daily operations.

## *5.3 Environmental challenges*

In order to facilitate the implementation of technology in the airport, the cooperation of multiple stakeholders (airlines, airport operators, regulatory agencies, etc), is required for the solution to be feasible, efficient and desirable for use by passengers (*96*). For example, an ideal scenario is the existence of an universal protocol for all airports in the adoption of RFID tags for baggage tagging (*71*). This provides a common database for easy retrieval of data and implementation in both departure and destination airports.

Furthermore, passengers need to be receptive towards technology adoption in airport operations and processes. A survey conducted revealed that passengers perceive information obtained through their own input or interpretation such as through map reading or a search on an airport wayfinding application to be less reliable, unofficial and more cognitive demanding as compared to information presented in "official" wall signs or from airport staff (*98*). This lack of trust towards SSTs can be attributed to the lack of technological expertise or the need for interaction particularly among older passengers (*105*). This presents a challenge for technologies that encourage passenger self-service such as airport mobile applications and self-help information kiosks as airports may need to adjust these technology offerings to cater to different demographics and passenger profiles.

## 6. Charting the way forward for Airport 4.0

From the SLR, it is evident that the implementation of I4.0 technologies is slowly gaining traction within the airport environment, and will continue to remain relevant in the digital transformation journeys in developing future airports. However, as I4.0 technologies have been typically used in non-airport (e.g. manufacturing) contexts, academic researchers have yet to explore the strategic use of digital technologies in the context of an airport and analyse their impact on customer experience (*34*). This contributed to the loose adoption of the term "Airport 4.0" in literature to suggest the ongoing digitalisation trends in airports in line with I4.0, with minimal references to constituent technologies or a proper definition of the phenomenon. Thus, future research can delve into a proper definition of the term "Airport 4.0" to describe the desired state of a smart, digital airport of the future and the enabling technologies to achieve this desired state.

Second, the adoption and implementation of a technology or innovation is a strategic multi-stage process (*106*). From the review, the implementation of technology in airports are mainly application-based and adopted in silos, instead of being part of an overall digital IT strategy. There is also a lack of quantitative assessment methods to evaluate airport technologies and build reliable business cases relating to new technology implementation (*43*). Considering the increasing trend of digital technology adoption in airports, the need for further research in digital or technology adoption strategy is necessary (*34*). This can be supplemented by a set of generic criteria to assess the technical features of the Airport 4.0 technologies (e.g. reliability, and lifetime) and the economic case for implementation (e.g. cost reductions, and passenger satisfaction).

Third, the challenges of adopting I4.0 technologies in airports were sparsely discussed in literature and majority of these challenges were identified by researchers rather than by airport operators. Future research can look into obtaining primary data from airport operators to verify the challenges they face in adopting I4.0 technologies as well as consolidate any best practices that support technology adoption in airports. Considering that airports vary in terms of size, layout, governance structure and other factors, developing a readiness assessment framework based on the challenges and best practices identified can allow airport operators to assess their current status and the problems they face before determining their suitability to adopt certain technologies. This readiness assessment can form part of an overall technology adoption strategy for airports to develop a roadmap with guidelines on infrastructural or process provisions that need to be in place as they move towards Airport 4.0.

Lastly, the papers reviewed mainly focused on the implementation of technologies for day-to-day operations to improve operational efficiency and enhance passenger experiences, with cost pressures being the main driver for digitalization (*43*). Less attention was focused on technology adoption to tackle uncertain and random events in the future such as crises or disasters which airports have little control of, but yet present a real test to an airport's resilience. These events include but are not limited to flight delays, technical disruptions, cybersecurity, aviation accidents, pandemics, economic downturns, etc. A handful of papers cover the use of data analytics to predict events such as flight delays and poor weather conditions, but many have also highlighted the lack of complete, quality datasets that can be used for prediction. Thus, future research can explore how digitalisation can help airports to better manage exceptional situations (crises and disasters) aside from solving day-to-day operational issues.

Despite the challenges identified in technology adoption within the unique airport context, airports should continue to harness technologies for airport operations. The ongoing COVID-19 pandemic has only exacerbated the need for airports to explore contactless technologies and robotics among other technologies that are closely related to I4.0.

**References**

1. M. Mira, M. H. Jeridi, D. Djabour, T. Ezzedine, Optimization of IoT Routing Based on Machine Learning Techniques: Case Study of Passenger Flow Control in Airport 3.0. 136–141 (2018), doi: 10.1109/IINTEC.2018.8695294

2. ICAO, Guidance for Air Travel through the COVID-19 Public Health Crisis, https://www.icao.int/covid/cart/Pages/CART-Take-off.aspx (2020).

3. S. E. Zaharia, C. V. Pietreanu, Challenges in airport digital transformation. *Transportation Research Procedia*. **35**, 90–99 (2018), doi:10.1016/j.trpro.2018.12.016

4. E. Buzzoni, F. Forlani, C. Giannelli, M. Mazzotti, S. Parisotto, A. Pomponio, C. Stefanelli, The advent of the internet of things in airfield lightning systems: Paving the way from a legacy environment to an open world. *Sensors (Switzerland)*. **19** (2019), doi:10.3390/s19214724.

5. S. Nielsen, E. Bonnerup, A. K. Hansen, J. Nilsson, L. J. Nellemann, K. D. Hansen, D. Hammcrshøi, Subjective Experience of Interacting with a Social Robot at a Danish Airport. 1163–1170 (2018), doi:10.1109/ROMAN.2018.8525643

6. V. Alcácer, V. Cruz-Machado, Scanning the Industry 4.0: A Literature Review on Technologies for Manufacturing Systems. *Engineering Science and Technology, an International Journal*. **22**, 899–919 (2019), doi:10.1016/j.jestch.2019.01.006

7. S. Wang, J. Wan, D. Zhang, D. Li, C. Zhang, Towards smart factory for industry 4.0: A self-organized multi-agent system with big data based feedback and coordination. *Computer Networks*. **101**, 158–168 (2016), doi:10.1016/j.comnet.2015.12.017

8. H. Lasi, P. Fettke, H.-G. Kemper, T. Feld, M. Hoffmann, Industry 4.0. *Business and Information Systems Engineering*. **6**, 239–242 (2014), doi:10.1007/s12599-014-0334-4

9. J. Lee, H.-A. Kao, S. Yang, Service innovation and smart analytics for Industry 4.0 and big data environment. **16**, 3–8 (2014), doi:10.1016/j.procir.2014.02.001

10. Y. Lu, Industry 4.0: A survey on technologies, applications and open research issues. *Journal of Industrial Information Integration*. **6**, 1–10 (2017), doi:10.1016/j.jii.2017.04.005

11. L. Thames, D. Schaefer, Software-defined Cloud Manufacturing for Industry 4.0. *Procedia CIRP*. **52**, 12–17 (2016), doi:10.1016/j.procir.2016.07.041.

12. A. A. Malik, T. Masood, A. Bilberg, Virtual reality in manufacturing: immersive and collaborative artificial-reality in design of human-robot workspace. *International Journal of Computer Integrated Manufacturing*. **33**, 22–37 (2020), doi:10.1016/j.jmsy.2020.09.008

13. T. Masood, J. Egger, M. Kern, Future-proofing the Through-life Engineering Service Systems. *Procedia Manufacturing*. **16**, 179–186 (2018), doi:10.1016/j.promfg.2018.10.162

14. A. A. Malik, T. Masood, R. Kousar, Repurposing factories with robotics in the face of COVID-19. *Science Robotics,* **3** (2020), doi:10.1126/scirobotics.abc2782

15. A. R. McLean, Coming to an Airport Near You. *Science*. **342**, 1330–1331 (2013), doi:10.1126/science.1247830

16. J. Mou, C. Liu, S. Chen, G. Huang, X. Lu, Temporal Characteristics of the Chinese Aviation Network and their Effects on the Spread of Infectious Diseases. *Sci Rep*. **7**, 1275 (2017). doi:10.1038/s41598-017-01380-5

17. S. Vandermerwe, J. Rada, Servitization of business: Adding value by adding services. *European Management Journal*. **6**, 314–324 (1988), doi:10.1016/0263-2373(88)90033-3

18. D. C. Fettermann, C. G. S. Cavalcante, T. D. de Almeida, G. L. Tortorella, How does Industry 4.0 contribute to operations management? *Journal of Industrial and Production Engineering*. **35**, 255–268 (2018), doi:10.1080/21681015.2018.1462863

19. T. Masood, P. Sonntag, Industry 4.0: Adoption challenges and benefits for SMEs. *Computers in Industry*, 26 (2020) doi:10.1016/j.compind.2020.103261

20. D. Tranfield, D. Denyer, P. Smart, Towards a Methodology for Developing Evidence-Informed Management Knowledge by Means of Systematic Review. *British Journal of Management*. **14**, 207–222 (2003), doi:10.1111/1467-8551.00375

21. R. B. Briner, D. Denyer, in *The Oxford Handbook of Evidence-Based Management* (2012).

22. R. Felkel, D. Steinmann, F. Follert, in *Digital Marketplaces Unleashed* (2017), pp. 443–453.

23. F. Koenig, P. A. Found, M. Kumar, Innovative airport 4.0 condition-based maintenance system for baggage handling DCV systems. *International Journal of Productivity and Performance Management*. **68**, 561–577 (2019), doi:10.1108/IJPPM-04-2018-0136

24. N. Barera, L. Sylvain, G. Valléry, S. Sutter, Conditions of use and adoption of digital tools: Results from a field study on airport reception agents, *Advances in Intelligent Systems and Computing*. **822** (2019), doi: 10.1007/978-3-319-96077-7_57

25. R. Almashari, G. Aljurbua, L. Alhoshan, N. S. Al Saud, O. Binsaeed, N. Nasser, IoT-based Smart Airport Solution (2018), doi:10.1109/SMARTNETS.2018.8707393

26. K. Straker, C. Wrigley, Engaging passengers across digital channels: An international study of 100 airports. *Journal of Hospitality and Tourism Management*. **34**, 82–92 (2018), doi:10.1016/j.jhtm.2018.01.001

27. A. M. Hainen, S. M. Remias, D. M. Bullock, Collection and analysis of multi-modal airport land side probe data from Bluetooth enabled mobile devices. 1304–1309 (2013), doi:10.1109/ITSC.2013.6728411

28. W. Huang, Y. Lin, B. Lin, L. Zhao, Modeling and predicting the occupancy in a China hub airport terminal using Wi-Fi data. *Energy and Buildings*. **203** (2019), doi:10.1016/j.enbuild.2019.109439

29. M. Joosse, V. Evers, A guide robot at the airport: First impressions. 149–150 (2017), doi:10.1145/3029798.3038389

30. R. Triebel, K. Arras, R. Alami, L. Beyer, S. Breuers, R. Chatila, M. Chetouani, D. Cremers, V. Evers, M. Fiore, H. Hung, O. A. I. Ramírez, M. Joosse, H. Khambhaita, T. Kucner, B. Leibe, A. J. Lilienthal, T. Linder, M. Lohse, M. Magnusson, B. Okal, L. Palmieri, U. Rafi, M. Van Rooij, L. Zhang, SPENCER: A socially aware service robot for passenger guidance and help in busy airports, *Springer Tracts in Advanced Robotics*, **113** (2016), doi:10.1007/978-3-319-27702-8_40

31. H. Sadjadi, M. A. Jarrah, Autonomous cleaning system for Dubai International Airport. *Journal of the Franklin Institute*. **348**, 112–124 (2011), doi:10.1016/j.jfranklin.2009.02.015

32. H. K. Yau, H. Y. H. Tang, Analyzing customer satisfaction in self-service technology adopted in airports. *Journal of Marketing Analytics*. **6**, 6–18 (2018), doi:10.1057/s41270-017-0026-2

33. V. Bogicevic, M. Bujisic, A. Bilgihan, W. Yang, C. Cobanoglu, The impact of traveler-focused airport technology on traveler satisfaction. *Technological Forecasting and Social Change*. **123**, 351–361 (2017). doi:10.1016/j.techfore.2017.03.038

34. S. Jaffer, G. Timbrell, Digital strategy in airports, paper presented at 25th Australasian conference on information system (2014).

35. R. Neufville, A. Odoni, *Airport Systems Planning, Design and Management* (ed. 2nd, 2013).

36. L. Hockley, Coronavirus roundtable: How is the aviation industry responding to the COVID-19 pandemic? *International Airport Review*, https://www.internationalairportreview.com/article/114585/aviation-industry-covid-19-pandemic/ (2020)

37. Schüller K., Wrobel C. (2018) Unlocking the Doors of Frankfurt Airport's Digital Marketplace: How Fraport's Smart Data Lab Manages to Create Value from Data and to Change the Airport's Way of Thinking. In: Linnhoff-Popien C., Schneider R., Zaddach M. (eds) Digital Marketplaces Unleashed. Springer, Berlin, Heidelberg. doi:10.1007/978-3-662-49275-8_57

38. C. M. Ariyawansa and A. C. Aponso, "Review on state of art data mining and machine learning techniques for intelligent Airport systems," *2016 2nd International Conference on Information Management (ICIM)*. 134–138 (2016), doi: 10.1109/INFOMAN.2016.7477547.

39. S. Mouli, AIRPORTS COUNCIL INTERNATIONAL, 16.

40. CAG, Bigger and better than before, https://www.changiairport.com/corporate/media-centre/resources/publication/issue-17/bigger-and-better-than-before-ishopchangi.html (2018).

41. P. Adhi, T. Burns, A. Davis, S. Lal, B. Mutell, "A transformation in store", *McKinsey & Company*, https://www.mckinsey.com/business-functions/operations/our-insights/a-transformation-in-store (2019).

42. IAR, Airports' IT spending to top $4.6 billion by 2023, new analysis claims. *International Airport Review*, https://www.internationalairportreview.com/news/70113/airports-spending-top-4-billion/ (2018).

43. I. Kovynyov, R. Mikut, Digital technologies in airport ground operations. *NETNOMICS: Economic Research and Electronic Networking* (2019), doi:10.1007/s11066-019-09132-5.

44. C. S. Leung, W.-D. Hao, C. M. Montiel, Piezoelectric sensors for taxiway airport traffic control system. 134–141 (2013), doi:10.1109/SusTech.2013.6617310

45. S. Futatsumori, K. Morioka, A. Kohmura, K. Okada, N. Yonemoto, Experimental feasibility study of 96 GHz FMCW millimeter-wave radar based upon radio-over-fiber technology-fundamental radar reflector detection test on the Sendai airport surface. 235–236 (2014), doi:10.1109/MWP.2014.6994540

46. J. Zhang, W. Chen, *Design of a monitoring system of airport boarding bridge based on ZigBee wireless network*. 2486–2491 (2014). doi:10.1109/CCDC.2014.6852591

47. J. Mund, A. Zouhar, L. Meyer, H. Fricke, C. Rother, Performance evaluation of LiDAR point clouds towards automated FOD detection on airport apron. 85–94 (2015), doi:10.1145/2899361.2899370

48. T. Martelli, C. Bongioanni, F. Colone, P. Lombardo, L. Testa, A. Meta, Security enhancement in small private airports through active and passive radar sensor (2016), doi:10.1109/IRS.2016.7497360

49. Garcia J., Pirovano A., Royer M. (2018) Routing in Wireless Sensor Networks for Surveillance of Airport Surface Area. In: Moreno García-Loygorri J., Pérez-Yuste A., Briso C., Berbineau M., Pirovano A., Mendizábal J. (eds) Communication Technologies for Vehicles. Nets4Cars/Nets4Trains/Nets4Aircraft 2018. Lecture Notes in Computer Science, vol 10796. Springer, Cham. doi:10.1007/978-3-319-90371-2_3

50. F. Donadio, J. Frejaville, S. Larnier, S. Vetault, Artificial intelligence and collaborative robot to improve airport operations, *Networks and Systems*. **22** (2018), doi:10.1007/978-3-319-64352-6_91.

51. A. Singh, M. Melo, Sensor design and layout for airport asphalt pavement instrumentation to test for delamination. 710–711, paper presented at Industrial Engineering and Operations Management (2017).

52. Z. Dong, X. Ma, X. Shao, Airport pavement responses obtained from wireless sensing network upon digital signal processing. *International Journal of Pavement Engineering*. **19**, 381–390 (2018), 10.1080/10298436.2017.1402601

53. M. Weiszer, J. Chen, P. Stewart, A real-time Active Routing approach via a database for airport surface movement. *Transportation Research Part C: Emerging Technologies*. **58**, 127–145 (2015), doi:10.1016/j.trc.2015.07.011

54. H. Fan, P. K. Tarun, V. C. P. Chen, D. T. Shih, J. M. Rosenberger, S. B. Kim, R. A. Horton, Data-driven optimization for Dallas Fort Worth International Airport deicing activities. *Annals of Operations Research*. **263**, 361–384 (2018), doi:10.1007/s10479-017-2747-1

55. H. Khadilkar, H. Balakrishnan, B. Reilly, Analysis of airport performance using surface surveillance data: A case study of BOS (2011), doi:10.2514/6.2011-6986

56. S. Bijjahalli, S. Ramasamy, R. Sabatini, A Novel Vehicle-Based GNSS Integrity Augmentation System for Autonomous Airport Surface Operations. *Journal of Intelligent and Robotic Systems: Theory and Applications*. **87**, 379–403 (2017), doi:10.1007/s10846-017-0479-8

57. S. El-Ansary, O. M. Shehata, E.-S. I. Morgan, Airport management controller: A multi-robot task-allocation approach. 26–30 (2016), doi:10.1145/3029610.3029623

58. S. Frank, P. M. Schachtebeck, P. Hecker, Sensor concept for highly-automated airport tugs for reduced emisson taxi operations, paper presented at 30[th] ICAS Congress (2016).

59. S. Conversy, J. Garcia, G. Buisan, M. Cousy, M. Poirier, N. Saporito, D. Taurino, G. Frau, J. Debattista, Vizir: A domain-specific graphical language for authoring and operating airport automations. 261–273 (2018), doi:10.1145/3242587.3242623.

60. G. B. França, M. V. de Almeida, S. M. Bonnet, F. L. Albuquerque Neto, Nowcasting model of low wind profile based on neural network using SODAR data at Guarulhos Airport, Brazil. *International Journal of Remote Sensing*. **39**, 2506–2517 (2018), doi:10.1080/01431161.2018.1425562

61. G. Zazzaro, G. Romano, P. Mercogliano, V. Rillo, S. Kauczok, Short range fog forecasting by applying data mining techniques: Three different temporal resolution models for fog nowcasting on CDG airport. 448–453 (2015), doi:10.1109/MetroAeroSpace.2015.7180699

62. F. Keis, WHITE - Winter hazards in terminal environment: An automated nowcasting system for Munich Airport. *Meteorologische Zeitschrift*. **24**, 61–82 (2014), doi:10.1127/metz/2014/0651

63. V. Ramanujam, H. Balakrishnan, Data-Driven Modeling of the Airport Configuration Selection Process. *IEEE Transactions on Human-Machine Systems*. **45**, 490–499 (2015), doi:10.1109/THMS.2015.2411743

64. H. Lee, W. Malik, Y. C. Jung, Taxi-out time prediction for departures at charlotte airport using machine learning techniques. 1–11 (2016), doi:10.2514/6.2016-3910

65. G. Xiangmin, M. Li, Departure capacity prediction for hub airport in thunderstorm based on data mining method. 6004–6009 (2017), doi:10.1109/CCDC.2017.7978246

66. Q. Wu, *A stochastic characterization based data mining implementation for airport arrival and departure delay data*, Applied Mechanics and Materials. 668–669 (2014). doi:10.4028/www.scientific.net/AMM.668-669.1037

67. J. Roudet, P.-E. Thurat, N. Turcot, Airport Ground-traffic Surveillance Systems Data Feed Innovative Comprehensive Analysis. **14**, 3741–3750 (2016), doi:10.1016/j.trpro.2016.05.459

68. J. Li, X. Wang, Y. Xu, Q. Li, C. He, Y. Li, Analyzing impact factors of airport taxiing delay based on ADS-B data. **42**, 1251–1255 (2017), doi:10.5194/isprs-archives-XLII-2-W7-1251-2017

69. X. Luo, Y. Tang, H. Wu, D. He, A New Method for Automatically Labeling Aircrafts in Airport Video. **22** (2015), doi:10.1051/matecconf/20152205010

70. W. Le, Application of Wireless Sensor Network and RFID Monitoring System in Airport Logistics. *Int. J. Onl. Eng.* **14**, 89 (2018), doi:10.3991/ijoe.v14i01.8058

71. W. Shehieb, H. Al Sayed, M. M. Akil, M. Turkman, M. A. Sarraj, M. Mir, A smart system to minimize mishandled luggage at airports. 154–158 (2017), doi:10.1109/PIC.2016.7949485

72. Y. Rouchdi, A. Haibi, K. El Yassini, M. Boulmalf, K. Oufaska, paper presented in *IEEE 5th International Congress on Information Science and Technology (CiSt),* 642–649 (2018)

73. M. Ghazal, S. Ali, F. Haneefa, A. Sweleh, Towards smart wearable real-time airport luggage tracking (2016), doi:10.1109/ICCSII.2016.7462422

74. W.-J. Park, D. H. Shin, S. Woo, S. Lee, Identification of applications of mobile devices to improve airport BHS maintenance. *KSCE Journal of Civil Engineering*. **18**, 1207–1212 (2014), doi:10.1007/s12205-014-0080-7

75. O. Koseoglu, E. T. Nurtan-Gunes, Mobile BIM implementation and lean interaction on construction site: A case study of a complex airport project. *Engineering, Construction and Architectural Management*. **25**, 1298–1321 (2018), doi:10.1108/ECAM-08-2017-0188

76. O. Koseoglu, B. Keskin, B. Ozorhon, Challenges and enablers in BIM-enabled digital transformation in mega projects: The Istanbul new airport project case study. *Buildings*. **9** (2019), doi:10.3390/buildings9050115

77. T. D. Oesterreich, F. Teuteberg, Understanding the implications of digitisation and automation in the context of Industry 4.0: A triangulation approach and elements of a research agenda for the construction industry. *Computers in Industry*. **83**, 121–139 (2016), doi:10.1016/j.compind.2016.09.006

78. G. Zhuo, paper presented in 2018 2nd IEEE Advanced Information Management, Communicates, Electronic and Automation Control Conference (IMCEC), 2479–2482 (2018).

79. O. Szymanezyk, T. Duckett, P. Dickinson, Agent-based crowd simulation in airports using games technology, *Computer Science (including subseries Lecture Notes in Artificial Intelligence and Lecture Notes in Bioinformatics)*, **7430** (2012), doi:10.1007/978-3-642-34645-3_9.

80. B. Baspinar, N. K. Ure, E. Koyuncu, G. Inalhan, Analysis of Delay Characteristics of European Air Traffic through a Data-Driven Airport-Centric Queuing Network Model. *IFAC-PapersOnLine*. **49**, 359–364 (2016), doi:10.1016/j.ifacol.2016.07.060.

81. R. An, X. K. Yang, *Application of simulation technology on airport system*, *Applied Mechanics and Materials*, **392** (2013), doi:10.4028/www.scientific.net/AMM.392.936

82. M. M. Mota, P. Scala, G. Boosten, Simulation-based capacity analysis for a future airport. 97–101 (2014), doi:10.1109/APCASE.2014.6924479

83. D. Gillen, W. G. Morrison, Aviation security: Costing, pricing, finance and performance. *Journal of Air Transport Management*. **48**, 1–12 (2015), doi:10.1016/j.jairtraman.2014.12.005

84. Y. Sterchi, A. Schwaninger, paper presented in 2015 International Carnahan Conference on Security Technology (ICCST*)*, 55–60 (2015).

85. N. Hättenschwiler, Y. Sterchi, M. Mendes, A. Schwaninger, Automation in airport security X-ray screening of cabin baggage: Examining benefits and possible implementations of automated explosives detection. *Applied Ergonomics*. **72**, 58–68 (2018), doi:10.1016/j.apergo.2018.05.003

86. B. Elias, Airport body scanners: The role of advanced imaging technology in airline passenger screening, *Airport baggage and passenger screening: Technology Elements and Consideration.* 1–18 (2013).

87. A.-M. Oostveen, M. Kaufmann, E. Krempel, G. Grasemann, Automated border control: A comparative usability study at two European airports. 27–34 (2014), doi:10.2139/ssrn.2432461

88. D. Ortega del Campo, C. Conde, Á. Serrano, I. Martín de Diego, E. Cabello, Face Recognition-based Presentation Attack Detection in a Two-step Segregated Automated Border Control e-Gate - Results of a Pilot Experience at Adolfo Suárez Madrid-Barajas Airport, *Proceedings of the 14th International Joint Conference on e-Business and Telecommunications*. 129–138 (2017), doi:10.5220/0006426901290138

89. N. Zhang, H.-Y. Jeong, A retrieval algorithm for specific face images in airport surveillance multimedia videos on cloud computing platform. *Multimedia Tools and Applications*. **76**, 17129–17143 (2017), doi:10.1007/s11042-016-3640-7

90. A. Davis, H. Chang, Airport protection using wireless sensor networks. 36–42 (2012), doi:10.1109/THS.2012.6459823

91. M. Zyczkowski, M. Szustakowski, W. Ciurapiński, R. Dulski, M. Kastek, P. Trzaskawka, Integrated mobile radar-camera system n airport perimeter security (2011), doi:10.1117/12.900332.

92. H. Wu, Z. Wang, C. Wang, Study on the recognition method of airport perimeter intrusion incidents based on laser detection technology. *Turkish Journal of Electrical Engineering and Computer Sciences*. **25**, 2737–2748 (2017), doi:10.3906/elk-1603-248.

93. P. Falcone, C. Bongioanni, A. Macera, F. Colone, D. Pastina, P. Lombardo, E. Anniballi, R. Cardinali, Active and passive radar sensors for airport security. 314–321 (2012), doi:10.1109/TyWRRS.2012.6381148

94. C. A. Vertua, L. Saini, O. Baud, N. Honore, P. E. Lawrence, Multi-approach strategy for multi-sensor data fusion enhancement: Unified analysis of advanced and unconventional techniques, targeted to airport surveillance enhancement. 265–270 (2011).

95. M. Mullan, The data-driven airport: How daa created data and analytics capabilities to drive business growth, improve the passenger experience and deliver operational efficiency. *Journal of Airport Management*. **13**, 361–379 (2019).

96. N. A. R. Negri, G. M. R. Borille, V. A. Falcão, Acceptance of biometric technology in airport check-in. *Journal of Air Transport Management*. **81** (2019), doi:10.1016/j.jairtraman.2019.101720.

97. W. Ge, Y. Zhang, D. Shao, J. Wang, Characterizing Pudong International Airport Parking Dynamics by Using Smart Card Data. 2546–2557 (2015), doi:10.1061/9780784479292.234

98. T. Fei, N. De Joux, G. Kefalidou, M. D'Cruz, S. Sharples, Towards understanding information needs and user acceptance of mobile technologies to improve passenger experience in airports (2016), doi:10.1145/2970930.2970938

99. C. Merschbrock, C. Nordahl-Rolfsen, BIM technology acceptance among reinforcement workers - The case of oslo airport's terminal 2. *Journal of Information Technology in Construction*. **21**, 1–12 (2016).

100. S. C. A. Thomopoulos, S. Daveas, A. Danelakis, Automated real-time risk assessment for airport passengers using a deep learning architecture. **11018** (2019), doi:10.1117/12.2519857

101. M. Tonkin, J. Vitale, S. Herse, M.-A. Williams, W. Judge, X. Wang, Design Methodology for the UX of HRI: A Field Study of a Commercial Social Robot at an Airport. 407–415 (2018), doi:10.1145/3171221.3171270

102. J. Lundberg, M. Nylin, B. Josefsson, Challenges for research and innovation in design of digital ATM controller environments: An episode analysis of six simulated traffic situations at Arlanda airport (2016), doi:10.1109/DASC.2016.7777949

103. L. Masson, J. Guiochet, H. Waeselynck, A. Desfosses, M. Laval, Synthesis of safety rules for active monitoring: Application to an airport light measurement robot. 263–270 (2017), doi:10.1109/IRC.2017.11.

104. A. Di Vaio, L. Varriale, Blockchain technology in supply chain management for sustainable performance: Evidence from the airport industry. *International Journal of Information Management* (2019), doi:10.1016/j.ijinfomgt.2019.09.010.

105. J. I. Castillo-Manzano, L. López-Valpuesta, Check-in services and passenger behaviour: Self service technologies in airport systems. *Computers in Human Behavior*. **29**, 2431–2437 (2013), doi:10.1016/j.chb.2013.05.030.

106. F. Damanpour, S. Gopalakrishnan, Theories of organizational structure and innovation adoption: the role of environmental change, 24 (1998).

**Acknowledgments:** This work was conducted as part of the "Industrial Systems of the Future" research program at the University of Cambridge, Cambridge, UK. J.H. Tan acknowledges generous support of Changi Airport Group.

**Competing interests:** The authors have no competing interests.